\documentclass[final]{ias2}

\usepackage{graphicx} 
\usepackage{multirow}
\usepackage{array} 

\usepackage{hyperref} 
\begin{document}

\markboth{Net-proton measurements at RHIC and the QCD phase diagram}{Bedangadas Mohanty}

\title{Net-proton measurements at RHIC and the QCD phase diagram}

\author[ain]{Bedangadas Mohanty} 
\email{bedanga@niser.ac.in}
%\author[yain]{Pascal Quazi} 
%\email{pqr@yain.ac.in}
%\author[sin,ain]{Bellatrix \lowercase{de la} Cruz}
%\email{bdc@sin.ac.in}
%\address[sin]{Some Institute Name, Country - 400076, Earth}
\address[ain]{School of Physical Sciences, National Institute of
  Science Education and Research, Bhubaneswar - 751005, India}
%\address[yain]{Yet Another Institute Name, Country - 678342, Earth}

\begin{abstract}
Two measurements related to the proton and anti-proton production near
midrapidity in $\sqrt{s_{\rm {NN}}}$ = 7.7, 11.5, 19.6,
27, 39, 62.4 and 200 GeV Au+Au collisions using the STAR detector at
the Relativistic Heavy Ion Collider (RHIC) are discussed. At intermediate impact
parameters the slope parameter of the directed flow versus
rapidity  ($dv_{1}/dy$) for the net-protons shows a non-monotonic
variation as a function of the beam energy. This non-monotonic
variation is characterized by the presence of a minimum in
$dv_{1}/dy$ between $\sqrt{s_{\rm {NN}}}$ = 11.5 and 19.6 GeV and a
change in the sign of $dv_{1}/dy$ twice between $\sqrt{s_{\rm    {NN}}}$  = 7.7 and 39 GeV.
At small impact parameters the product of the moments of net-proton
distribution, kurtosis $\times$ variance ($\kappa\sigma^{2}$) and
skewness $\times$ standard deviation ($S\sigma$) are observed to be
significantly below the corresponding measurements at large impact
parameter collisions for $\sqrt{s_{\rm    {NN}}}$  = 19.6 and 27
GeV. The $\kappa\sigma^{2}$ and $S\sigma$ values at these beam
energies deviate from the expectations from Poisson statistics and
that from a Hadron Resonance Gas model.  Both these
measurements have implications towards the understanding of the QCD phase
structures, the first order phase transition and the critical point
in the high baryonic chemical potential region of the phase diagram.

\end{abstract}


 
\maketitle

% \tableofcontents
% \listoffigures
% \listoftables

\section{Introduction}

The formation of a hot and dense medium of deconfined quarks and
gluons (QGP) has been established in high energy heavy ion collisions at the
Relativistic Heavy Ion Collider (RHIC) facility at Brookhaven National
Laboratory and the Large Hadron Collider (LHC) facility at
CERN~\cite{starwhitepaper}. The
transition from QGP to a hadron gas has been shown to be a cross
over~\cite{Aoki:2006we}. The focus of research in this field has now
shifted towards two aspects,
(a) characterizing the transport properties of QGP and (b)
establishing the QCD phase structures at high baryonic chemical
potential ($\mu_{B}$) region of the QCD phase diagram. 

A rigorous phenomenological analysis of the precision data from
relativistic heavy-ion collisions and theoretical advances over 14
years has led to quantitative estimates for some of the transport
properties of a strongly interacting de-confined state of quarks and
gluons. The estimated shear viscosity to entropy density ratio
($\eta/s$) is found to reflect the inviscid liquid property of the QGP and
has a value between (1-2)/4$\pi$~\cite{Luzum:2008cw,Song:2010mg}. That reflecting the stopping power
or the opacity of QGP has been estimated by obtaining the square
of the momentum transferred by the parton to the QGP per unity length
($\hat{q}$) and is found to lie between 2-10 GeV$^2$/fm~\cite{Vitev:2002pf,Bass:2008rv}.

\begin{figure}
 \begin{center}
%\vspace{-3.0cm}
\includegraphics[scale=0.5]{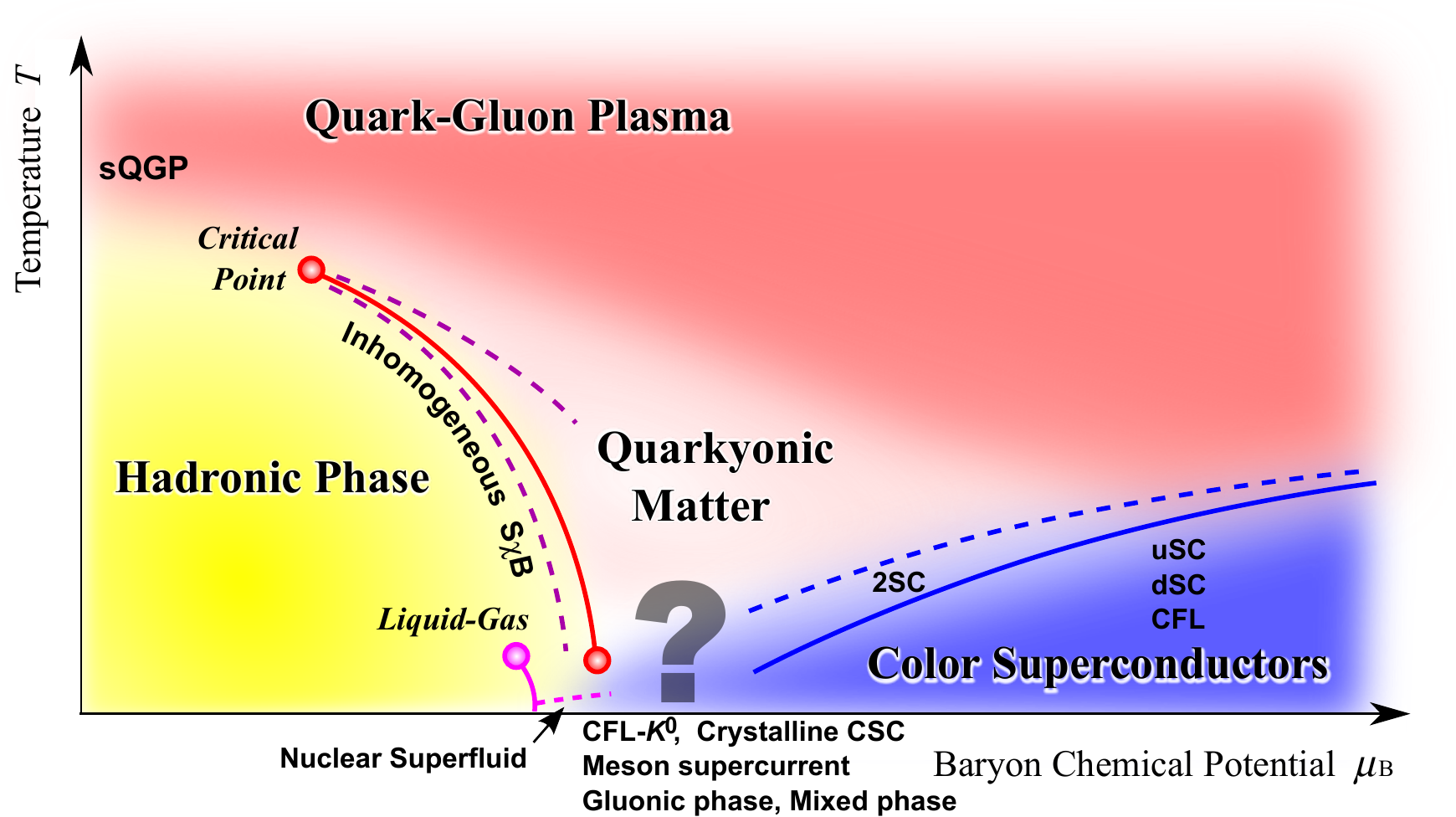}
 \end{center}
%\vspace{-4.0cm}
 \caption{Conjectured QCD phase diagram~\cite{Fukushima:2010bq} .}
 \label{phasediagram}
\end{figure}
On the other hand a dedicated program called the Beam Energy Scan
(BES) to establish the phase diagram
of QCD was launched at RHIC in the year 2010 to unravel the QCD phase
structure at large $\mu_{B}$. A range of $\mu_{B}$ from 20 MeV to 400
MeV of the phase diagram was covered by varying the $\sqrt{s_{NN}}$
from 200 to 7.7 GeV. The rich phase structure in the high $\mu_{B}$
region can be seen from the conjectured QCD phase diagram shown in
Fig.~\ref{phasediagram}~\cite{Fukushima:2010bq}. The two distinct features of the phase
diagram at large $\mu_{B}$ are the first order phase boundary and the
critical point (CP). In this paper we concentrate on the status of the experimental
search for these two phase structures through the measurement of
proton and antiproton production in heavy-ion
collisions. Specifically we discuss the observable related to the
azimuthal and multiplicity distributions for net-proton
(the difference in number of protons and anti-protons) in Au+Au 
collisions at midrapidity for $\sqrt{s_{\rm {NN}}}$ = 7.7, 11.5, 19.6,
27, 39, 62.4 and 200 GeV. We find the results to be very
intriguing, which must to be further quantified by having a high event
statistics second phase of BES program in near future at RHIC.

In the next section we discuss the two observables related to the
search for the first order phase transition and the critical point. In
section 3 we present the experimental results on the directed
flow measurements of net-protons, an observable for first order phase
transition. In section 4 we discuss the measurements related to the
product of higher moments of net-proton multiplicity distribution,
observable for critical point search. Finally in section 5 we summarize
the findings. 

\section{Observables}
The experimental results presented here are from the data recorded in
the STAR detector at RHIC in the year 2010 and 2011. 
%The proton and
%anti-proton are produced in the Au+Au collisions at $\sqrt{s_{NN}}$ = 7.7,
%11.5, 19.6, 27, 39, 62.4 and 200 GeV.

\subsection{Directed flow}
The patterns of azimuthal anisotropy in particle production, often
termed as flow, in heavy-ion collisions can be obtained by studying
the Fourier expansion  of the azimuthal angle ($\phi$) distribution of
produced particles with respect to the reaction plane angle
($\Psi_{R}$)~\cite{oai:arXiv.org:nucl-ex/9805001}. 
Where the reaction plane is defined as the plane subtended by the
impact parameter direction and the beam direction. 
The various (order $n$) coefficients in this expansion are defined as:
\begin{equation}
v_{n}=\langle\cos[n(\phi-\Psi_{R})]\rangle.
\end{equation}
The angular brackets in the definition denote an average over 
many particles and events. Directed flow is quantified by the first coefficient ($v_{1}$). On
the other hand the elliptic flow is given by the second coefficient
($v_{2}$).

The $v_{1}$, which is sensitive to early collision dynamics, is
proposed as a signature of first order phase transition based on a
hydrodynamic calculation~\cite{oai:arXiv.org:nucl-th/9505014,
  oai:arXiv.org:nucl-th/9908010, oai:arXiv.org:nucl-th/0406018}.
These calculations whose equation of state incorporates a first-order phase transition
from hadronic matter to QGP predict a non-monotonic variation of the
slope of directed flow of baryons (and net-baryons) around midrapidity as a function of
beam energy, also has a prominent minimum, and a double sign change in the $v_1$ slope, which is not seen in the
same hydrodynamic model without a first-order phase transition.  

The $v_{1}$ results discussed in this paper are for the most abundantly
measured baryons, antiproton and proton, detected using their energy loss in STAR Time Projection
Chamber and by the time-of-flight information from the Time Of Flight
detector~\cite{Adamczyk:2014ipa}. Protons and antiprotons have a transverse momentum between
0.4 and 2.0 GeV/$c$ and pseudorapidity ($\eta$)  between $\pm$ 1 unit.  The first
order event plane for $\sqrt{s_{NN}}$  $<$ 62.4 GeV is constructed
using the information from the two Beam Beam Counters at  3.3 $<$ $\mid
\eta \mid$ $<$ 5.0. That for $\sqrt{s_{NN}}$  = 62.4 and 200 GeV uses
the information from STAR ZDC-SMD detectors. All results discussed
here are corrected for event plane resolution and proton track reconstruction
efficiency.

\subsection{Higher moments} 
Non-monotonic variations of observables related to the moments of the 
distributions of conserved quantities such as net-baryon, net-charge, 
and net-strangeness~\cite{volker} number with $\sqrt{s_{\mathrm {NN}}}$ are
believed to be good signatures of  a CP. 
The moments are related to the correlation length ($\xi$) of the system~\cite{stephanovmom}.
%The signatures of phase transition or CP are detectable if they 
%survive the evolution of the system~\cite{survival}. 
Finite size and  time effects in heavy-ion collisions put constraints 
%on the values of $\xi$ and hence 
on the significance of the desired signals. A theoretical calculation suggests 
a non-equilibrium $\xi$ $\approx$ 2-3 fm for heavy-ion collisions~\cite{krishnaxi}. Hence, it is
proposed to study the higher moments (like skewness, 
${\it {S}}$ = $\left\langle (\delta N)^3 \right\rangle/\sigma^{3}$ 
and kurtosis, $\kappa$ = [$\left\langle (\delta N)^4 \right\rangle/\sigma^{4}$] -- 3 
with $\delta N$ = $N$ -- $\langle N \rangle$) of distributions of conserved quantities
due to a stronger dependence on $\xi$~\cite{stephanovmom}. 
%Both the magnitude
%and the sign of the moments~\cite{asakawa}, which quantify the shape of the multiplicity
%distributions, are important for understanding phase transition and CP effects.
Further, products of the moments can be related to susceptibilities associated 
with the conserved numbers. The product $\kappa$$\sigma^2$ of the net-baryon number
distribution is related to the ratio of fourth order ($\chi^{(4)}_{\mathrm B}$) 
to second order ($\chi^{(2)}_{\mathrm B}$) baryon number susceptibilities~\cite{latticesus,Gavai:2010zn}. 
The ratio $\chi^{(4)}_{\mathrm B}$/$\chi^{(2)}_{\mathrm B}$ is expected to deviate
from unity near the CP. It has different values for the hadronic and
partonic phases~\cite{Gavai:2010zn}. 

The higher moments of the net-proton multiplicity
($N_{p} - N_{\bar{p}}$ = $\Delta N_{p}$) distributions from Au+Au
collisions discussed in this paper are for protons and anti-protons
detected at midrapidity ($|y|$ $<$ 0.5) in the range 
0.4 $<$ $p_{\mathrm T}$ $<$ 0.8 GeV/$c$.  A good purity of the
proton sample (better than 98\%) for all beam energies is obtained in
the momentum range studied. All results presented are corrected for
proton reconstruction efficiency. 

\section{Search for the first order phase transition}
\begin{figure}
\begin{center}
\includegraphics[scale=0.45]{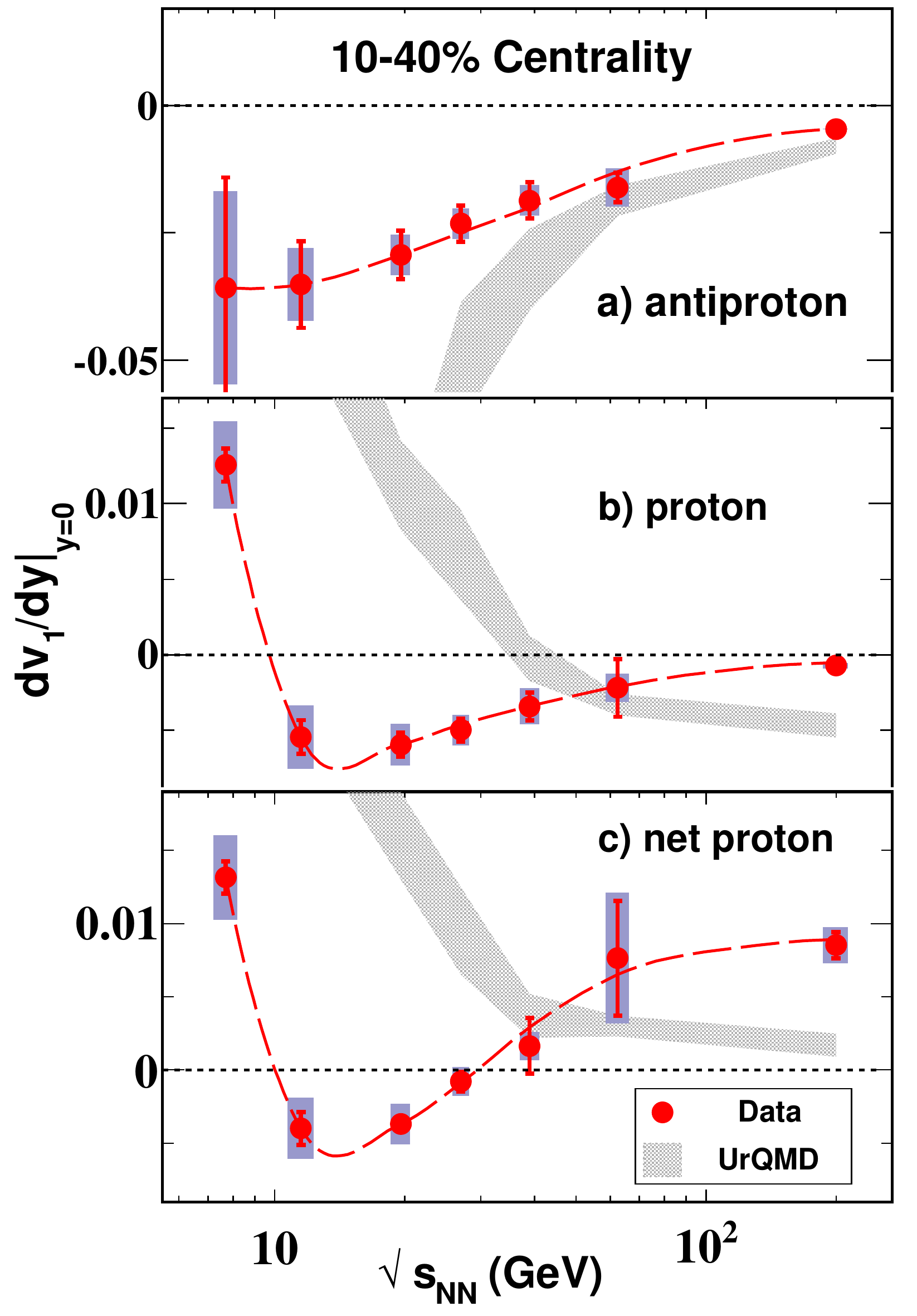}
\caption{(Color online) Directed flow slope ($dv_1/dy$) near mid-rapidity as a
  function of beam energy ($\sqrt{s_{NN}}$) for
  intermediate-centrality (10-40\%) Au+Au collisions~\cite{Adamczyk:2014ipa}.  Panels (a), (b)
  and (c) shows the STAR experiment measurement for antiprotons,
  protons, and net-protons, respectively, along with corresponding
  calculations from the UrQMD model obtained with  the same cuts and
  fit conditions.  The systematic uncertainties on the measurements
  are shown as shaded bars. The dashed curves are a smooth fit to
  guide  the eye. }
\label{fig:v1}
\end{center}
\end{figure} 
Figure~\ref{fig:v1} shows the slope of directed flow versus rapidity
($dv_{1}/dy$) near midrapidity as a function of $\sqrt{s_{NN}}$ for
antiprotons (panel (a)), protons (panel (b)) and net-proton (panel
(c))~\cite{Adamczyk:2014ipa}. The results are for 10-40\% Au+Au collision centrality. The
antiproton $dv_{1}/dy$ has negative values and shows a monotonic
increase with  $\sqrt{s_{NN}}$. The proton $dv_{1}/dy$ has positive
values for $\sqrt{s_{NN}}$ = 7.7 GeV and negative values for rest of
the energies studied. The proton $dv_{1}/dy$ dependence on $\sqrt{s_{NN}}$ is
non-monotonic with a minimum around $\sqrt{s_{NN}}$ = 11.5 and 19.6
GeV. The energy dependence of proton $dv_1/dy$ involves an interplay
between the directed flow of protons associated with baryon number
transported from the initial beam rapidity to the vicinity of
mid-rapidity, and the directed flow of protons from
particle-antiparticle pairs produced near mid-rapidity.  The
importance of the pair production mechanism increases strongly with
beam energy. It is important to distinguish between the two mechanisms
before further conclusions can be drawn. The net-proton $dv_1/dy$ is
expected to provide the contribution from protons associated with
baryon number transport. Assuming antiproton directed flow as a proxy
for the directed flow of pair produced protons, the proposed
net-proton slope  can be constructed from  $$[v_1(y)]_p = r(y) [v_1(y)]_{\bar{p}} + [1-r(y)]\, [v_1(y)]_{{\rm net\mbox{-}}p}\,,$$ 
where $r(y)$ is the observed rapidity dependence of the ratio of
antiprotons to  protons at each beam
energy~\cite{Adamczyk:2014ipa}. The net-proton slope is shown as a
function of $\sqrt{s_{NN}}$ in Fig.~\ref{fig:v1}(c). The data shows a
non-monotonic dependence on  $\sqrt{s_{NN}}$ with a minimum around
$\sqrt{s_{NN}}$ = 11.5 and 19.6 GeV. The values of slope changes sign
twice, goes from positive at 7.7 GeV to negative at $\sqrt{s_{NN}}$  =
11.5 - 27 GeV, then again becomes positive for $\sqrt{s_{NN}}$ $>$ 39
GeV. 

The corresponding UrQMD
results~\cite{oai:arXiv.org:nucl-th/9803035,oai:arXiv.org:hep-ph/9909407},
which does not include any first
order phase transition effects,  in all the three cases of slope of
antiproton, proton and net-proton shows a monotonic variation with
$\sqrt{s_{NN}}$. The slope values also does not agree with the
measurements. 

A possible interpretation of the changing sign of the $v_1$ slope is that it reflects a change
in equation-of-state. At a given energy where the system undertakes a first order quark--hadron
phase transition, one expect the formation of a mixed-phase for which the pressure
gradient is small. The softest pressure will naturally produces the observed minimum in
$v_1$ slope parameter~~\cite{oai:arXiv.org:nucl-th/9505014,
  oai:arXiv.org:nucl-th/9908010, oai:arXiv.org:nucl-th/0406018}.  The
alternate proposal is that at higher energies pair production is
dominant at mid-rapidity and transported baryons have relatively small influence. As there
is no preferred direction for pair produced hadrons, so the slope parameter becomes close
to zero. While at lower beam energies the observed baryons are strongly influenced by the
transported baryons so they are aligned with them hence the slope
parameter is positive. The most intriguing is the dependence in the
intermediate energies, a mean field model study shows that the energy
dependent  baryon potential plays an important role in this region~\cite{oai:arXiv.org:nucl-th/0502058}. 

In the search for the signature of a first-order phase transition in
the high $\mu_{B}$ region of the QCD phase diagram, the findings from
$dv_{1}/dy$ from the RHIC beam energy scan program are very compelling 
and strongly motivates further measurements. To better understand
the possible role and relevance of stopping in interpretation of existing data on net-proton
directed flow, new higher event statistics measurements of the
centrality dependence of $v_1$ at $\sqrt{s_{NN}}$  = 7.7 to 19.6 GeV is needed. 

\section{Search for the critical point}
\begin{figure}
\begin{center}
\includegraphics[scale=0.45]{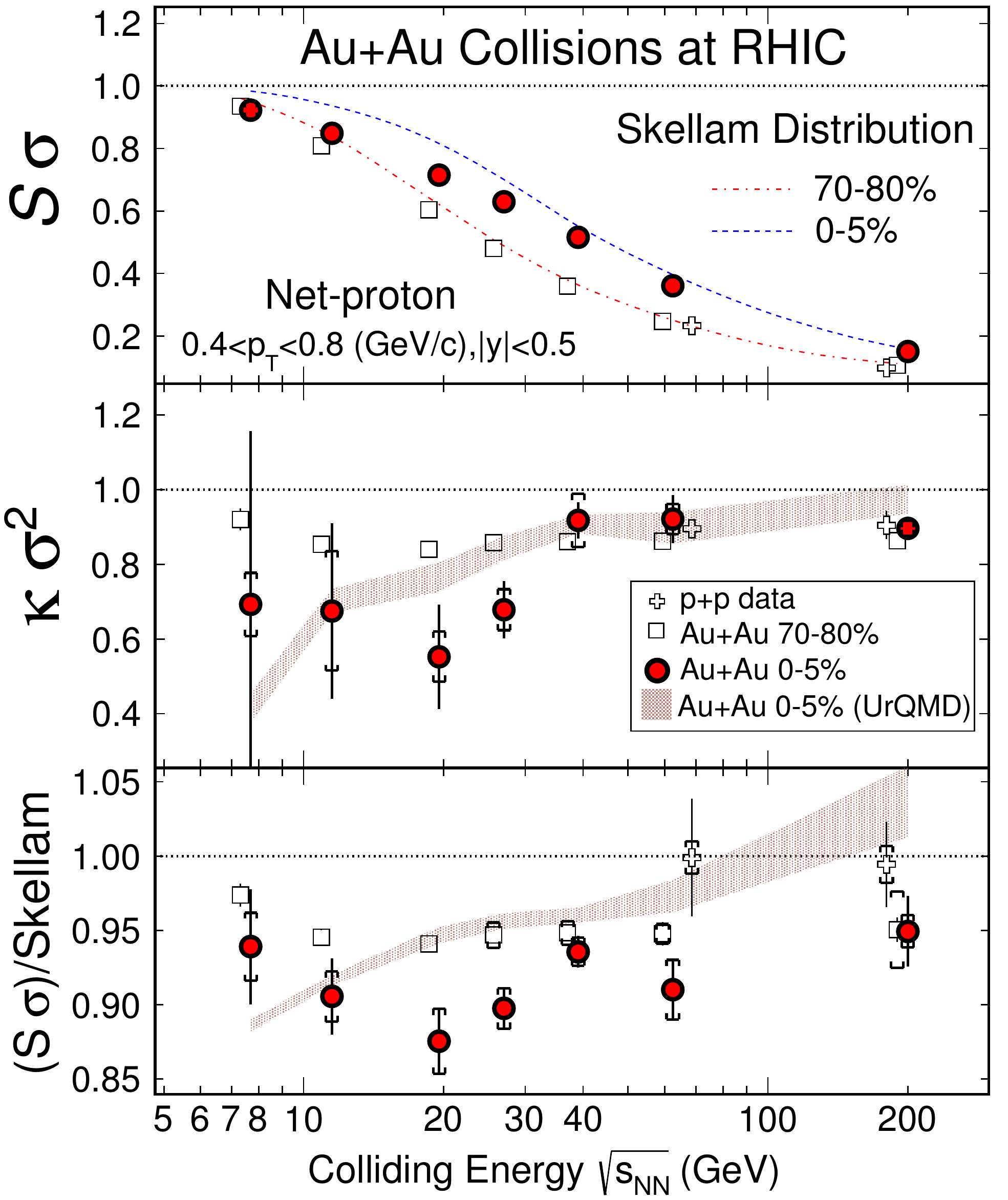}
\caption{(Color online) Collision energy and centrality dependence of
  the net-proton ${\it{S}}\sigma$ and $\kappa$$\sigma^2$
from Au+Au and $p$+$p$ collisions at RHIC~\cite{oai:arXiv.org:1309.5681}. 
Crosses, open squares and filled circles are for the efficiency
corrected results of $p$+$p$, 70-80\%,
and 0-5\% Au+Au collisions, respectively. Skellam distributions for corresponding 
collision centralities are shown in the top panel. Shaded hatched bands are the results from
UrQMD.  The hadron resonance gas model (HRG) values for $\kappa$$\sigma^2$ and ${\it{S}}\sigma$/Skellam are unity. 
The error bars are statistical and caps are systematic errors. 
For clarity, $p$+$p$ and 70-80\% Au+Au results are slightly displaced horizontally. }
\label{fig:cp}
\end{center}
\end{figure} 

Figure~\ref{fig:cp} shows the energy dependence of ${\it{S}}\sigma$ and $\kappa$$\sigma^2$ for $\Delta N_{p}$ 
for Au+Au collisions for two collision centralities (0-5\% and 70-80\%), corrected for $p$($\bar{p}$) reconstruction 
efficiency~\cite{STAR_netproton1, oai:arXiv.org:1309.5681}.  The
Skellam expectations for  ${\it{S}}\sigma$ are calculated using the
data as $(\langle N_{p} \rangle - \langle N_{\bar p} \rangle)/(\langle
N_{p} \rangle + \langle N_{\bar p} \rangle)$. 
The ${\it{S}}\sigma$ values normalized to the corresponding Skellam expectations are shown in
the bottom panel of Fig.~\ref{fig:cp}.  The Skellam expectations
reflect a system of totally uncorrelated, statistically random
particle production.  The central collision data shows deviation from  Skellam
expectation with maximum deviation occurring for $\sqrt{s_{\mathrm
    {NN}}}$ = 19.6 and 27 GeV. The corresponding results from $p$+$p$ collisions at
$\sqrt{s_{\mathrm {NN}}}$ = 62.4 and 200 GeV  are also shown and found to be similar to peripheral Au+Au 
collisions within the statistical errors. For $\sqrt{s_{\mathrm
    {NN}}}$ = 19.6 and 27 GeV, differences are observed between  the 0-5\% central Au+Au collisions and 
the peripheral collisions. The results are closer to unity for $\sqrt{s_{\mathrm {NN}}}$ = 7.7 GeV. 
Higher statistics data for $\sqrt{s_{\mathrm {NN}}}$ $<$ 19.6 GeV will
help in quantitatively understanding the suggestive non-monotonic energy
dependence of $\kappa$$\sigma^2$ and ${\it{S}}\sigma$.

The data also show deviations from the hadron resonance gas model~\cite{Karsch:2010ck,Garg:2013ata}
which predict $\kappa$$\sigma^2$ and  ${\it{S}}\sigma$/Skellam to be
unity. The effect of decay is less than 2\% as per the HRG calculations in Ref.~\cite{Garg:2013ata}. 
To understand the effects of baryon number conservation~\cite{Bzdak:2012an} and experimental acceptance,
UrQMD model calculations (a transport model which does not include a
CP)~\cite{oai:arXiv.org:nucl-th/9803035,oai:arXiv.org:hep-ph/9909407}  for 0-5\% Au+Au collisions 
are shown in the middle  and bottom panels of Fig.~\ref{fig:cp}. The
UrQMD model shows a monotonic decrease with decreasing beam energy~\cite{Luo:2013bmi}.
The centrality dependence of the $\kappa$$\sigma^2$ and
${\it{S}}\sigma$ from UrQMD~\cite{Luo:2013bmi} (not shown in the
figures) closely follow the data at the lower beam energies of 7.7 and
11.5 GeV. Their values are in general larger compared to data for the higher beam energies. 
The observed  $\sqrt{s_{\mathrm {NN}}}$ dependence for central Au+Au collisions is not explained by UrQMD model.

The current data provide the most relevant measurements over the widest range in $\mu_{B}$ (20 to 450 MeV) 
to date for the CP search, and for comparison with the baryon number
susceptibilities computed from QCD  to understand the various features 
of the QCD phase structure~\cite{latticesus,Gavai:2010zn}.
The deviations of ${\it{S}}$$\sigma$ and $\kappa$$\sigma^2$ below Skellam expectation are qualitatively
consistent with a QCD based model which includes a CP~\cite{Stephanov:2011zz}.
However, conclusions on the existence of CP can be made only after a
high event statistics measurement for $\sqrt{s_{NN}}$ $<$ 27 GeV in the
second phase of the beam energy scan program and comparison to QCD calculations with 
CP behavior.

\section{Summary}

We have discussed two striking observations from the RHIC beam energy
scan program related to the first order quark--hadron phase transition
and the critical point.  The measurements uses the produced protons
and antiprotons in Au+Au collisions at midrapidity for $\sqrt{s_{NN}}$
= 7.7, 11.5, 19.6, 27, 39, 62.4 and 200 GeV. 

The slope of the directed flow of protons and net-protons in mid-central collisions (10-40\% centrality) at
midrapidity ($dv_{1}/dy$)  shows a clear non-monotonic variation with
respect to $\sqrt{s_{NN}}$ ($\mu_{B}$). The minimum value of
$dv_{1}/dy$ lies somewhere between $\sqrt{s_{NN}}$ ($\mu_{B}$)
 = 27 GeV(160 MeV)  to 11.5 GeV(315 MeV). The net-proton $dv_{1}/dy$
 changes sign twice in the beam energy range studied. This observable which is
 driven by the pressure gradients developed in the system is sensitive
 to first order phase transition effects. The energy dependence of the measured
 $dv_{1}/dy$ is consistent with a theoretical hydrodynamic model
 calculation with first order phase transition~\cite{oai:arXiv.org:nucl-th/9505014,
  oai:arXiv.org:nucl-th/9908010, oai:arXiv.org:nucl-th/0406018}.

Deviations of $\kappa$$\sigma^2$ and $S\sigma$ for net-proton
distribution in 0-5\% centrality is observed at $\sqrt{s_{\mathrm {NN}}}$ = 19.6 and
27 GeV from: (a) 70-80\% peripheral collisions, (b) Poisson and hadron
resonance gas expectation value of close to unity and (c) transport
model based UrQMD calculation within the experimental acceptance. 
The deviations of ${\it{S}}$$\sigma$ and $\kappa$$\sigma^2$ below Poisson expectation are qualitatively
consistent with a QCD based model which includes a CP~\cite{Stephanov:2011zz}.
Higher statistics data set at $\sqrt{s_{NN}}$ $<$ 20 GeV in the second phase of the beam energy scan
program is needed to clarify whether the energy dependence of the observable will follow a non-monotonic
 variation with a minimum around $\sqrt{s_{NN}}$
 = 27 GeV to 11.5 GeV, as observed for the net-proton
 $dv_{1}/dy$, or a monotonic variation with $\sqrt{s_{NN}}$. 

\indent {\bf Acknowledgement}: This work is supported by the DST
Swarnajayanti Fellowship of Government of India.

%\bibliographystyle{pramana}
%\bibliography{references}

\end{document}